\documentclass[useAMS,usenatbib]{mn2e}
\usepackage{natbibmnfix}
\usepackage{graphicx}
\usepackage{grffile}
\usepackage{subfigure}
\usepackage{mathrsfs,amssymb}
\usepackage{amsmath}
\usepackage{natbib}
\usepackage{color}
\usepackage{ulem}
\usepackage{url}
\usepackage{bm}
\usepackage{multirow}
\usepackage[flushleft]{threeparttable}
\usepackage{gensymb}

\newcommand{\araa}{\rm {ARA\&A}}             % Annual Review of Astron and Astrophys
\newcommand{\apj}{\rm {ApJ}}                 % Astrophysical Journal
\newcommand{\apjl}{\rm {ApJ}}                % Astrophysical Journal, Letters
\newcommand{\aap}{\rm {A\&A}}                % Astronomy and Astrophysics
\newcommand{\mnras}{\rm {MNRAS}}             % Monthly Notices of the RAS
\newcommand{\nat}{\rm {Nature}}              % Nature
\def\gtsima{$\; \buildrel > \over \sim \;$}
\def\ltsima{$\; \buildrel < \over \sim \;$}
\def\gsim{\lower.5ex\hbox{\gtsima}}
\def\lsim{\lower.5ex\hbox{\ltsima}}
\def\simgt{\lower.5ex\hbox{\gtsima}}
\def\simlt{\lower.5ex\hbox{\ltsima}}
\def\simpr{\lower.5ex\hbox{\prosima}}

 \newcommand*\oline[1]{%
  \vbox{%
    \hrule height 0.5pt%                  % Line above with certain width
    \kern0.25ex%                          % Distance between line and content
    \hbox{%
      \kern-0.1em%                        % Distance between content and left side of box, negative values for lines shorter than content
      \ifmmode#1\else\ensuremath{#1}\fi%  % The content, typeset in dependence of mode
      \kern-0.1em%                        % Distance between content and left side of box, negative values for lines shorter than content
    }% end of hbox
  }% end of vbox
}

\begin{document}

\title[NIRB fluctuations]{Updated analysis of near-infrared background fluctuations}
\author[Yue et al.]{Bin Yue$^{1}$, Andrea Ferrara$^{1, 2}$, Ruben Salvaterra$^{3}$  \\
 $^1$Scuola Normale Superiore, Piazza dei Cavalieri 7, I-56126 Pisa, Italy\\
 $^2$Kavli IPMU (WPI), Todai Institutes for Advanced Study, the University of Tokyo, Japan\\
 $^3$INAF/IASF-Milan, via E. Bassini 15, I-20133 Milano, Italy\\
 }

\maketitle

\begin{abstract}
The power spectrum of Near InfraRed Background (NIRB) fluctuations measured at 3.6 $\mu$m by  {\tt  Spitzer} shows a clustering excess over the known galaxies signal that has been interpreted in terms of early ($z\simgt 13$), accreting (direct collapse) black holes (DCBH) or low-$z$ intrahalo light (IHL). In addition, these fluctuations correlate with the cosmic X-ray background (CXB) measured at (0.5-2) keV, supporting the black hole explanation. This scenario has been questioned by the recent detection of a correlation between the two  {\tt CIBER} 1.1/1.6 $\mu$m bands with the 3.6 $\mu$m  {\tt  Spitzer} one. This correlation is hardly explained by early DCBHs that, due to intergalactic absorption, cannot contribute to the shortest wavelength bands. Here we show that the new correlation is caused instead by a Diffuse Galactic Light (DGL) component arising from Galactic stellar light scattered by dust. The black hole interpretation of the excess remains perfectly valid and, actually, the inclusion of DGL allows less demanding (by up to about 30\%) requirements on the DCBH abundance/mass.  
\end{abstract}


\section{Introduction}
On angular scales $\simgt 100$ arcsec (corresponding to multipole number $\ell \lsim 10^4$)  the power spectrum of the source-subtracted Near InfraRed Background (NIRB) fluctuations has an amplitude that exceeds by $\simgt 100$ times the signal expected from all galaxies below the detection limit and first stars.  The origin of such ``clustering excess" (i.e. due to correlations in the sources spatial distribution) has now represented a puzzle since its discovery (\citealt{2005Natur.438...45K,2007ApJ...654L...5K,2011ApJ...742..124M, 2012ApJ...753...63K, 2012Natur.490..514C, 2014Sci...346..732Z, 2015ApJ...807..140S, 2012ApJ...756...92C, 2013MNRAS.431..383Y}).  

Two scenarios have been proposed to interpret the clustering excess. The first advocates the contribution from intrahalo light (IHL), i.e. relatively old stars stripped from their parent galaxies following merging events.  These stars therefore reside in between dark matter halos and constitute a low-surface brightness haze around galaxies. The IHL is expected to come mostly from low redshifts ($1+z\lsim 1.5$) systems \citep{2012Natur.490..514C,2014Sci...346..732Z}. 
 
The second scenario is instead based on the presence of a class of early, highly obscured accreting black holes of intermediate mass ($\sim10^{4-6} M_\odot$) at $z\gsim 13$ \citep{2013MNRAS.433.1556Y,2014MNRAS.440.1263Y}.  As a suitable mechanism to produce such objects does exist -- the so called Direct Collapse Black Holes (DCBH, for a concise overview of the problem see \citealt{2014MNRAS.443.2410F}), and the interpretation of the supermassive black holes observed at $z=6$ seemingly requires massive seeds \citep{2011RvMA...23..189V}, such hypothesis seems particularly worth exploring.

Both scenarios successfully explain the observed clustering excess,
albeit with apparently demanding requirements. In fact, if the excess
is to be explained by intrahalo light, then a large fraction of the stars at low-$z$ must reside outside systems that we would normally classify as ``galaxies'' \citep{2014Sci...346..732Z}.
On the other hand, in the DCBH scenario the abundance of seed black holes produced until $z\sim13$ must represent a sizeable fraction of the estimated present-day black hole abundance, as deduced from local scaling relations \citep{2013ARA&A..51..511K} and recently revised by \citet{2015A&A...574L..10C}. However, it is important to outline that both scenarios are not in conflict with any known observational evidence. 

The two scenarios, however, differ strongly for what concerns the interpretation of the observed cross-correlation between 3.6 $\mu$m NIRB fluctuations and those measured at (0.5-2) keV in the cosmic X-ray background (CXB) by \citet{2013ApJ...769...68C}. While accreting black holes naturally produce X-ray emission, no obvious similar mechanism can be identified  for intrahalo stars, thus making difficult to explain the observed cross-correlation. 

The DCBH scenario might have its own problems. They could possibly arise from the recent measurement of NIRB fluctuations at 1.1 and 1.6 $\mu$m obtained by {\tt  CIBER}\footnote{\url{http://ciber.caltech.edu}}. \cite{2014Sci...346..732Z} showed that these fluctuations do correlate with those observed at 3.6 $\mu$m by {\tt Spitzer}, thus  suggesting a common source. If confirmed, this could possibly represent a problem for DCBHs as their formation must stop, based on a number of physical arguments discussed in \citet{2014MNRAS.440.1263Y}, after $z\simeq 13$. Intergalactic absorption at wavelengths shorter than the Ly$\alpha$ line would then prevent DCBH to contribute at  1.1 and 1.6 $\mu$m.

Here, we show that the cross-correlation between {\tt  CIBER}  and {\tt  Spitzer} bands does not represent a problem for the DCBH hypothesis, as very likely it arises from a well-known contaminant: Diffuse Galactic Light (DGL), i.e. dust scattered or thermally emitted light in the Milky Way\footnote{Often the term DGL specifically refers to stellar light scattered by dust, while longer wavelength ($\simgt 5 \mu$m) thermal emission from grains is referred to as the ``Galactic cirrus".}.  \cite{2014Sci...346..732Z} pointed out that DGL strongly contributes to the  {\tt CIBER} 1.1 and 1.6 $\mu$m auto-correlation power spectra on large scales. Motivated by this evidence, we show that the 1.1$\times$3.6  $\mu$m and 1.6$\times$3.6 $\mu$m cross-correlations at the largest scales ($\gsim 1\degree$) can be purely ascribed to the DGL and that
 there is no tension between the high redshift DCBH scenario and {\tt  CIBER} observations. Actually, the sub-dominant contribution of DGL in the 3.6$\mu$m band helps {\it decreasing} the DCBH abundance required to explain the clustering excess.

\section{methods}\label{methods}

In order to estimate the DGL component in the 3.6 $\mu$m auto-correlation power spectrum we proceed in the following
way. First, we use the fact that on small scales (and in all bands) shot noise completely dominates the 1-halo and 2-halo clustering terms both in the auto- and cross-correlation power spectra. Therefore we can safely assume that at these scales 
\begin{equation}
C_l\approx C_{\rm SN},
\end{equation} 
and derive the value of $C_{\rm SN}$ by fitting the five {\tt  CIBER} data points at the smallest scales in the 1.1 $\mu$m and 1.6 $\mu$m auto-correlation power spectra, and in the 1.1(1.6)$\times$3.6 $\mu$m cross-correlation ones. Results of the fit are reported in Table~\ref{fittings}.

Similarly, on the largest scales the signal is dominated by the DGL component, so that
\begin{equation}
C_l \approx C_{\rm DGL} = A_{\rm DGL} \left(\frac{l}{1000}\right)^{\alpha}.
\end{equation}
Here we assume, following \citet{2014Sci...346..732Z,2015NatCo...6E7945M}, that the DGL power spectrum has a power-law form with a fixed slope $\alpha = - 3.05$ for both the DGL auto-correlation and cross-correlation power spectra as derived by the combined fit of {\tt  CIBER} and {\tt HST} data\footnote{Note that the constraints are mainly provided by {\tt HST} data.} \citep{2015NatCo...6E7945M}.  Results of the fit of the six {\tt CIBER} data points on the largest scales are also given in Table~\ref{fittings}. We checked that our results do not change significantly when we let the value of $\alpha$ vary within its 1$\sigma$ errors  ($-3.05 \pm 0.07$).
\begin{table}
\begin{center}
\caption{Best fitting parameters for 1.1 and 1.6 $\mu$m auto-, and 1.1$\times$3.6, 1.6$\times$3.6 $\mu$m cross-correlation power spectra. We use the 5 (6) data points at the smallest (largest) angular scales for shot noise (DGL) component  fitting.}
\begin{tabular}{lcc}
\hline \hline
power spectrum & ${\rm log}(C_{\rm SN})$  &  ${\rm log}(A_{\rm DGL})$ \\
\hline
1.1$\times$1.1 $\mu$m&$-6.11^{+0.007}_{-0.008}$&  $-4.5^{+0.3}_{-1.7}$ \\
1.6$\times$1.6 $\mu$m&$-6.33^{+0.007}_{-0.01}$&$-4.6^{+0.3}_{-1.8}$   \\
1.1$\times$3.6 $\mu$m&$-7.69^{+0.01}_{-0.006}$& $-6.4^{+0.3}_{-1.1}$ \\
1.6$\times$3.6 $\mu$m&$-7.79^{+0.009}_{-0.008}$& $-6.4^{+0.3}_{-1.1}$\\
\hline
\hline
\end{tabular}\\
\label{fittings}
\end{center}
\end{table}
We further implicitly assume here that the DGL fluctuations at two wavelengths  are perfectly correlated (i.e. unity correlation coefficient). This seems a reasonable assumption given that the three bands very closely spaced in wavelength. At this point we can derive the contribution of DGL to the 3.6 $\mu$m auto-correlation power spectrum as
\begin{equation}
A_{\rm DGL}^{\rm ab} = \sqrt{ A_{\rm DGL}^{\rm aa}\times A_{\rm DGL}^{\rm bb}},
\end{equation}
which gives 
\begin{equation}\label{a3.6}
A_{\rm DGL}^{\rm bb}=(A_{\rm DGL}^{\rm ab})^2/A_{\rm DGL}^{\rm aa}.
\end{equation}
In the previous expressions, $b$ corresponds to the 3.6 $\mu$m band; $a$ corresponds either to the 1.1 or 1.6 $\mu$m band. 
\begin{figure*}
\centering{
\subfigure{\includegraphics[scale=0.4]{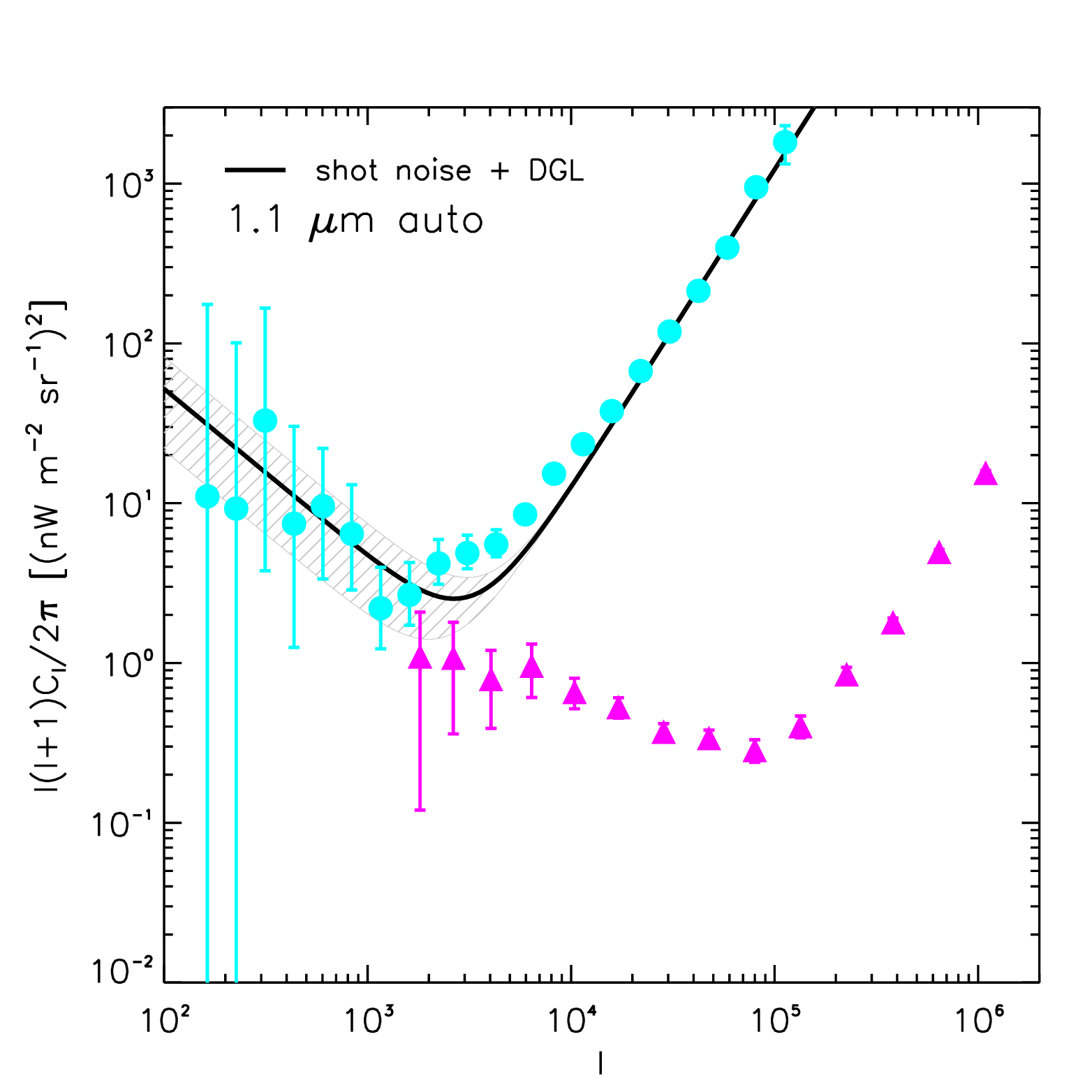}}
\subfigure{\includegraphics[scale=0.4]{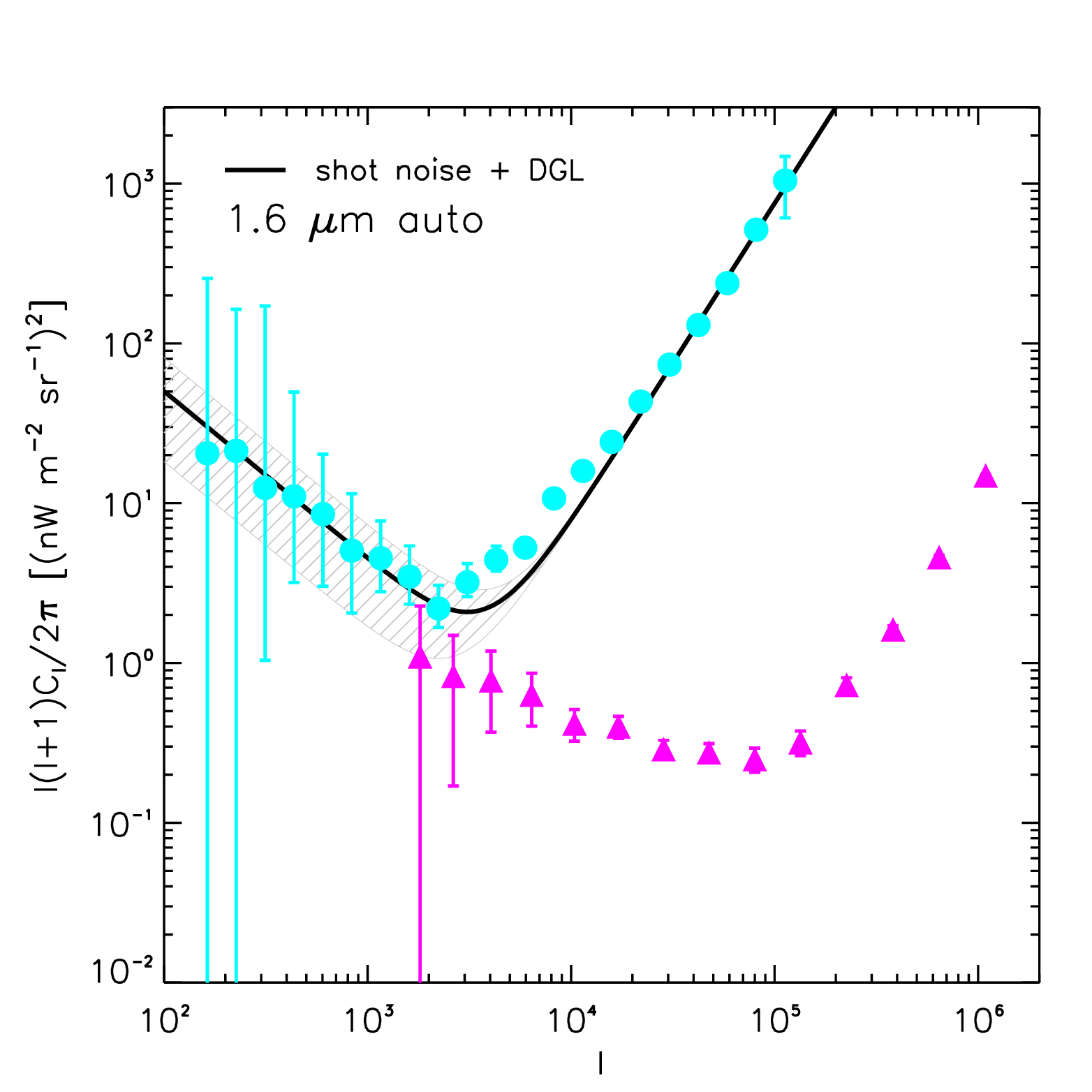}}
\subfigure{\includegraphics[scale=0.4]{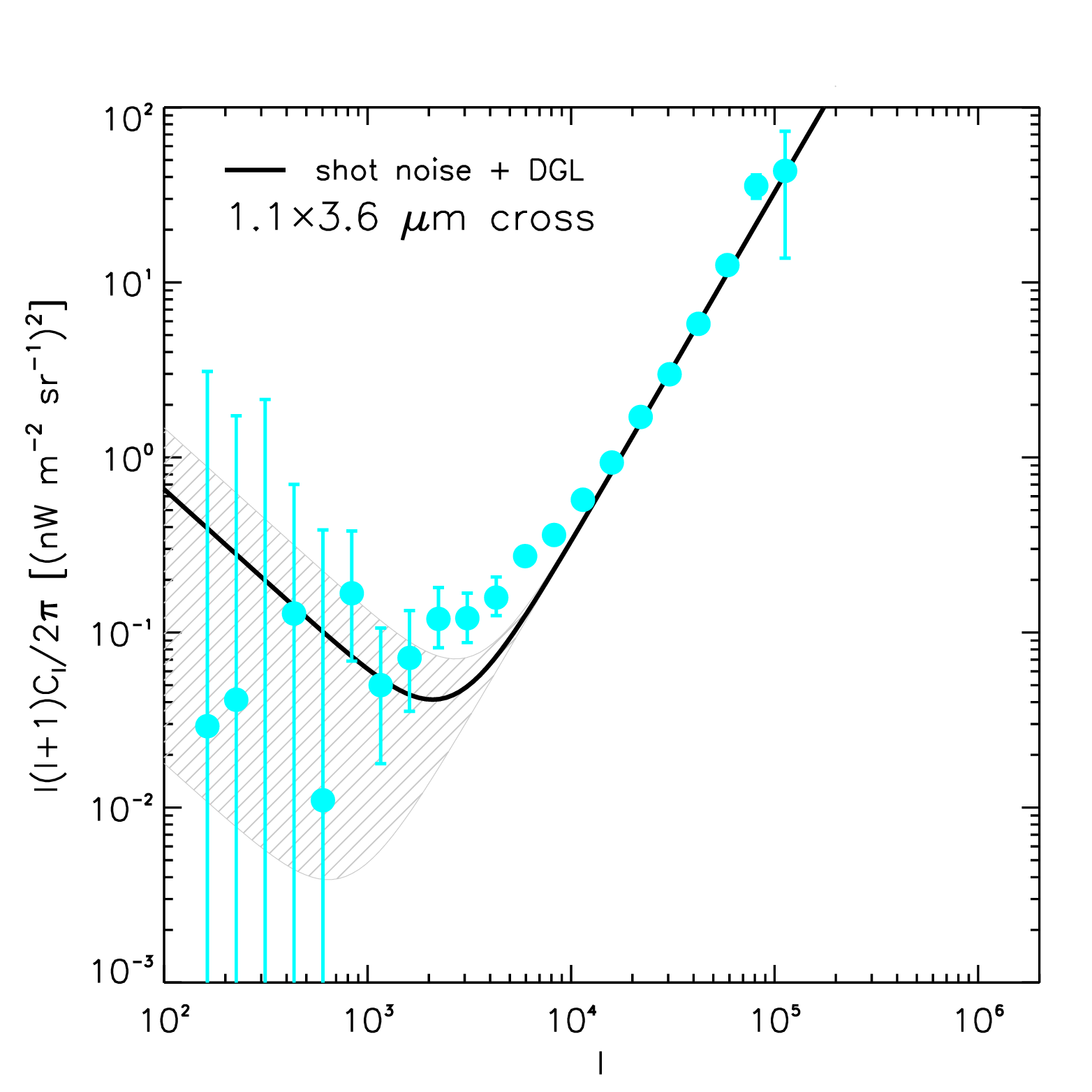}}
\subfigure{\includegraphics[scale=0.4]{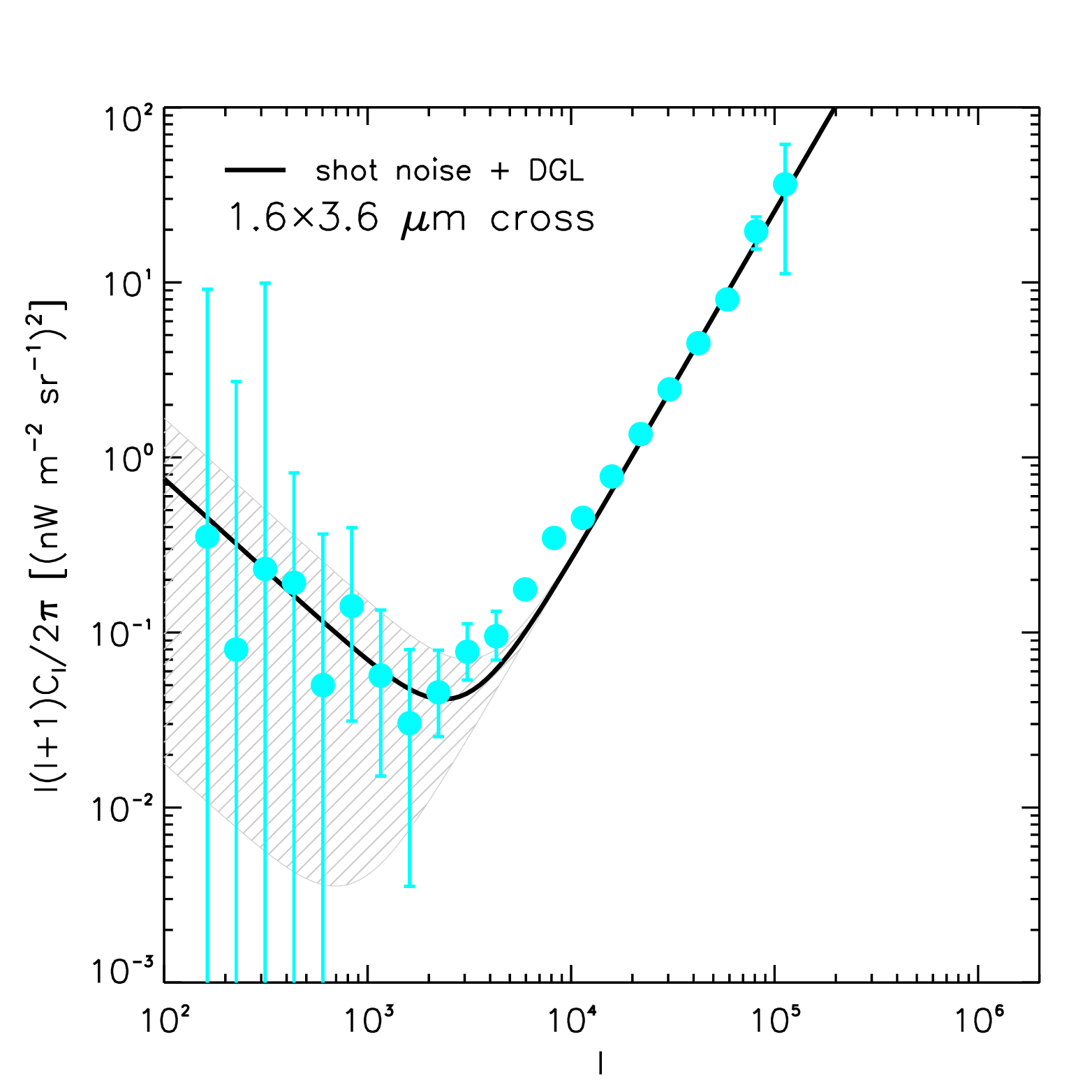}}
 \caption{{\it Upper:} The 1.1 (left) and 1.6 $\mu$m (right) {\tt CIBER} auto-correlation power spectra (filled circles); the solid line shows the sum of the best-fit shot noise and DGL contribution, with the 1$\sigma$ variance indicated by shaded areas.  Also plotted the HST power spectra from \citet{2015NatCo...6E7945M} (filled triangles); note that in the 1.1 $\mu$m panel the {\tt HST} points are actually measured at 1.25 $\mu$m. {\it Lower:} Same as above for the cross-correlation power
spectra. 
}
\label{auto}
}
\end{figure*}
\begin{figure*}
\centering{
\subfigure{\includegraphics[scale=0.4]{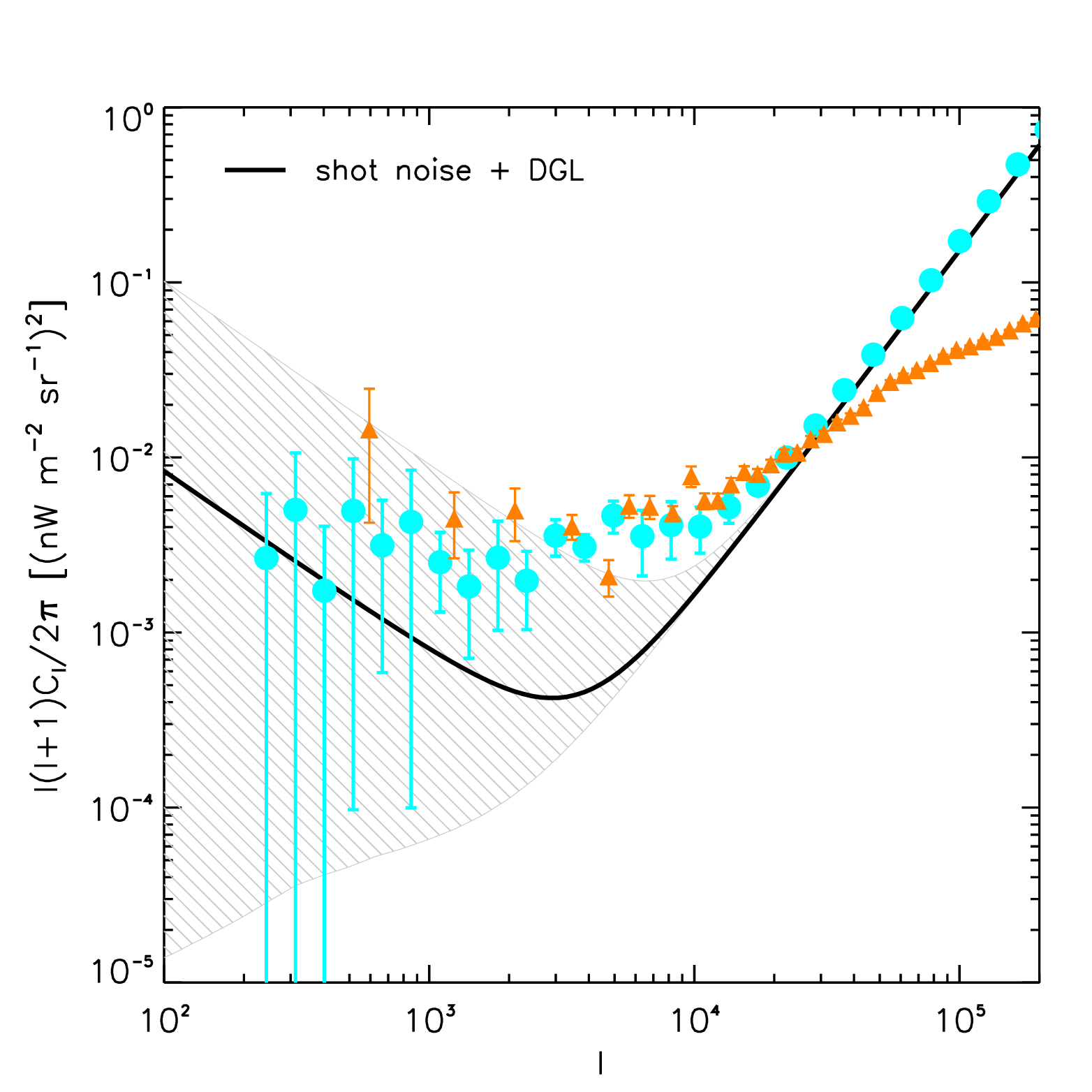}}
\subfigure{\includegraphics[scale=0.4]{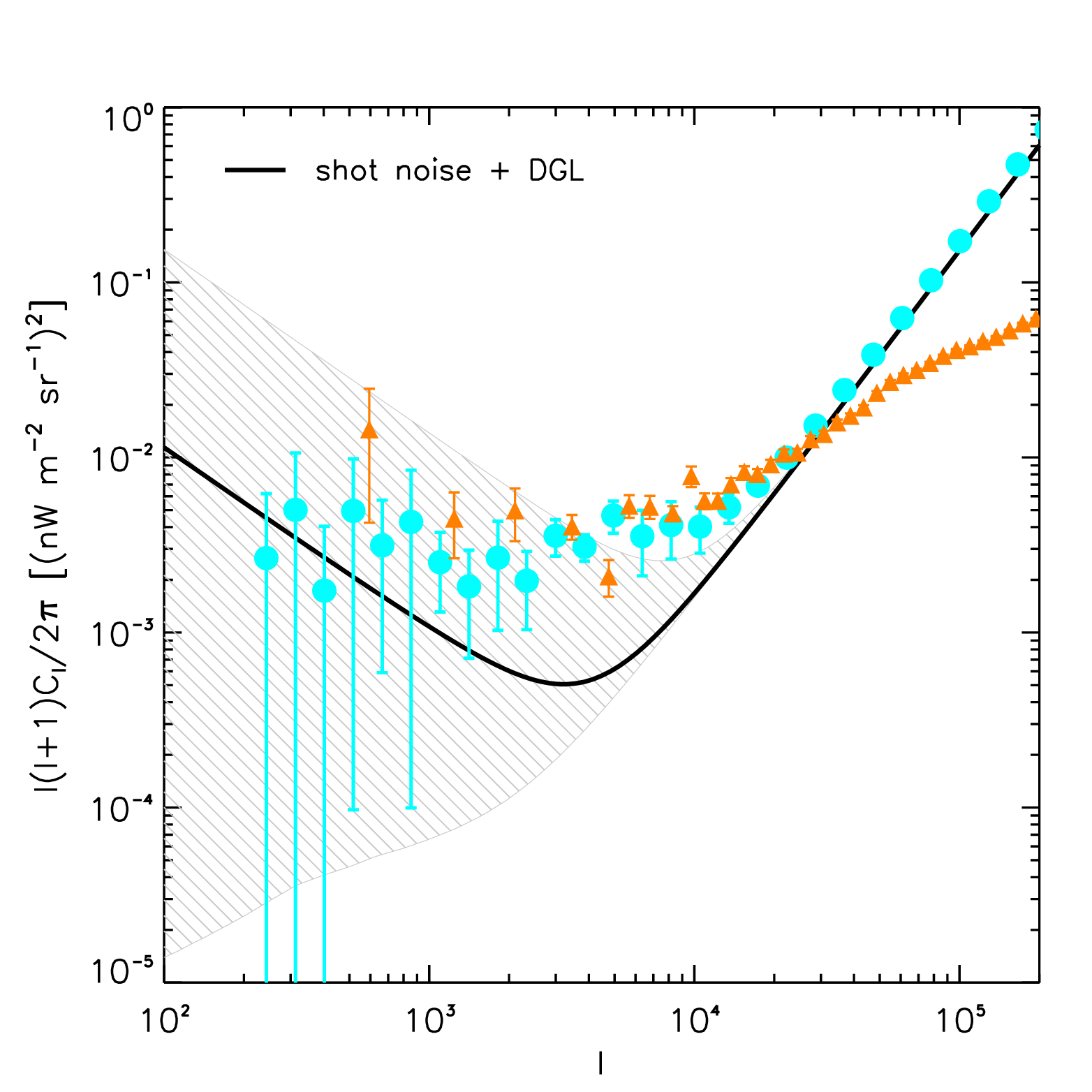}}
\subfigure{\includegraphics[scale=0.4]{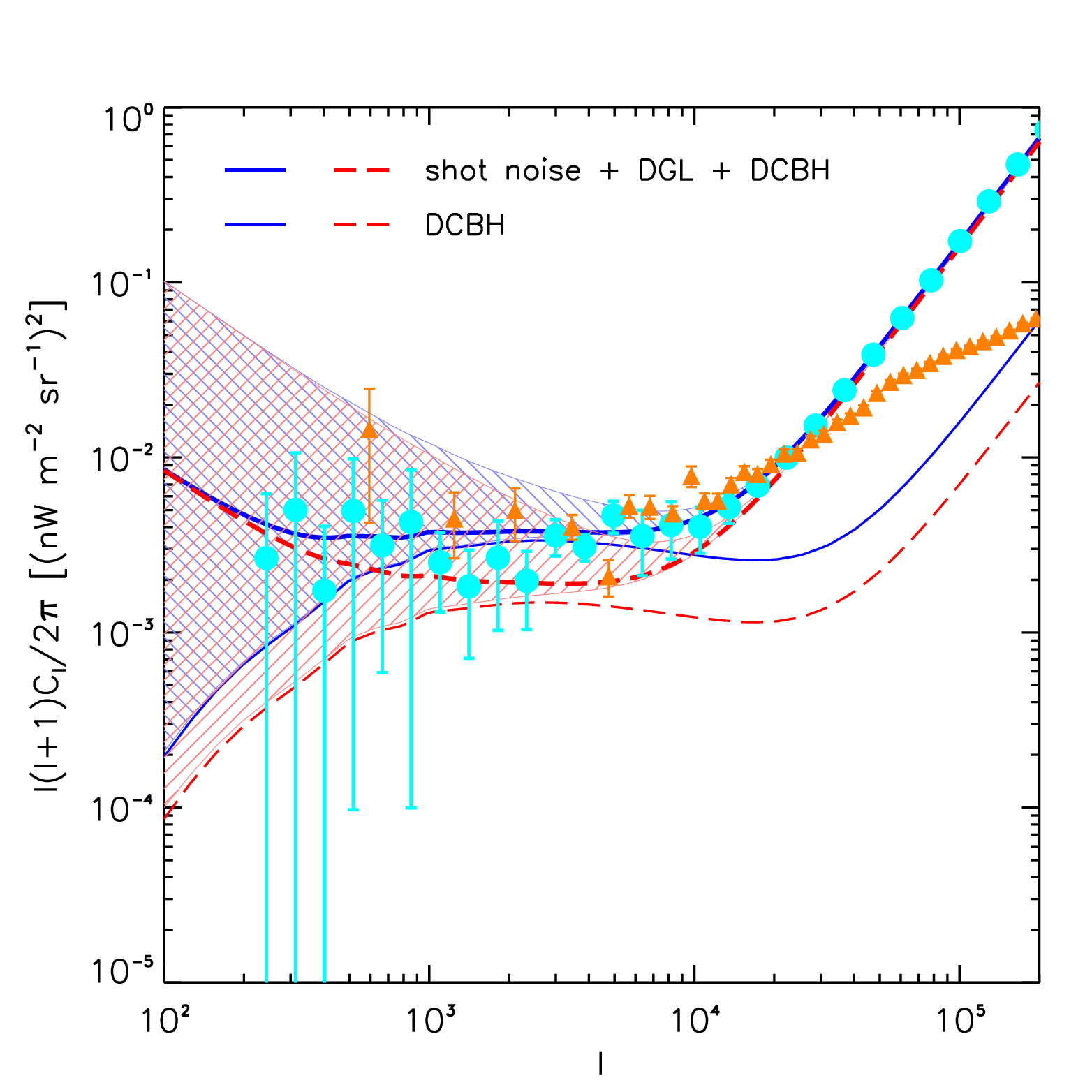}}
\subfigure{\includegraphics[scale=0.4]{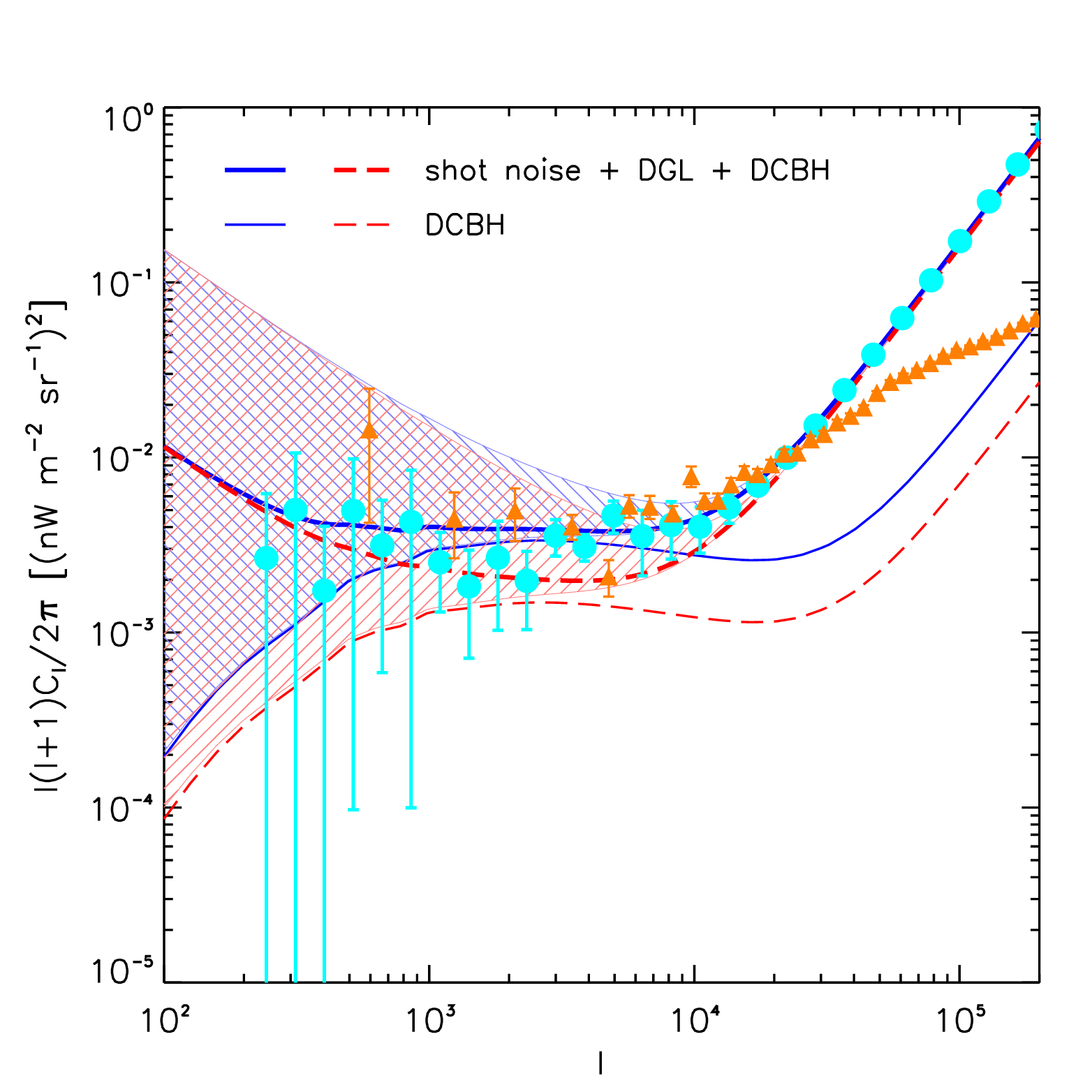}}
 \caption{The 3.6 $\mu$m auto-correlation power spectra derived from (1.1) 1.6 $\mu$m auto-power spectra and (1.1)1.6$\times$3.6 $\mu$m cross-correlation power spectra by assuming perfect correlations. \textit{Upper panels}: sum of
DGL and low-$z$ galaxies contributions.   \textit{Bottom}: we add the DCBH contribution (thin lines are for DCBH only, thick lines are the sum) consistent with a mass density of active DCBHs of $4\times10^5~M_\odot$Mpc$^{-3}$ at peak (solid curves and corresponding shaded regions) or $2.7 \times 10^5~M_\odot$ Mpc$^{-3}$  at peak (dashed curves with corresponding shaded regions). In each panel the filled circles (triangles) with error bars are data from \citet{2012Natur.490..514C} \citep{2012ApJ...753...63K}.
}
\label{auto36}
}
\end{figure*}
Note that the DGL is derived from the (1.1)1.6$\times$3.6 $\mu$m cross-correlation power spectra in \citet{2014Sci...346..732Z}. In this work {\tt Spitzer} observations were analyzed using a shallower source subtraction depth (mag $\simeq 22$) than in the original work by \citet{2012Natur.490..514C}, who instead used a deeper threshold (mag $\simeq 24$). As a result, the 3.6 $\mu$m auto-power spectrum in \citet{2014Sci...346..732Z} is largely dominated by shot noise and no interesting signal is seen. For this reason we have decided to compare the derived auto-power spectrum with the 3.6 $\mu$m auto-power in \citet{2012Natur.490..514C}. This should not introduce any artifacts, as the DGL component must be independent of the point source subtraction depth.  Also, note that the 
DGL fitting has been derived for the same fields, thus eliminating possible effects introduced by the dependence of the DGL signal from Galactic coordinates. 

\section{Results}\label{results}
We compare the curve $[A_{\rm DGL}(l/1000)^\alpha+C_{\rm  SN}]l^2/(2\pi)$ and its 1$\sigma$ variance with observations in
Fig. \ref{auto}. We find that both the {\tt CIBER} measured auto-correlation power spectra and the cross-correlation power spectra
could be nicely matched by the sum of shot noise and DGL without the need for additional components. 
Indeed, while some extra-power can be seen in {\tt CIBER} data at $3000 \lsim l \lsim 2\times10^4$, 
such an bump is not seen in \citet{2015NatCo...6E7945M} where resolved sources are subtracted down to a much deeper limiting
magnitude. Thus it is likely that this bump in {\tt CIBER} data can be due to foreground sources not resolved in \citet{2014Sci...346..732Z} analysis.

As described in the previous Section, we estimate the contribution of the DGL to the 3.6 $\mu$m auto-correlation power spectrum from both 1.1 $\mu$m and 1.6 $\mu$m data. We obtain a value for the DGL amplitude log($A_{\rm DGL})=-8.3_{-3.3}^{+1.1}$ and log($A_{\rm DGL})=-8.2_{-3.5}^{+1.1}$ from the 1.1 $\mu$m and 1.6 $\mu$m data, respectively.
In the top panels of  Fig. \ref{auto36} we show the 3.6 $\mu$m fluctuation spectrum measured by {\tt Spitzer} (\citealt{2012Natur.490..514C} for the same field, and \citealt{2012ApJ...753...63K} in a different field as comparison), along with the contribution from DGL and $z<5$ galaxies (clustering and shot noise) as computed by following the methods in \citet{2012ApJ...752..113H}. The solid line shows the sum of these two components while the light shaded area indicates the 1$\sigma$ uncertainty. 

Shot noise and DGL account quite well for the observed power on both large and small scales. However, the clustering excess is still clearly visible at intermediate scales, i.e. at $300\lsim l \lsim 2\times10^4$. In principle, such extra power might be partially accounted by the large errors associated to the estimated DGL component. However, the maximally allowed DGL power spectrum largely overestimates the observed signal at $l<2\times 10^3$. Moreover, we recall that if the 3.6 $\mu$m fluctuations were purely due to the combination of low-$z$ galaxies and DGL signals, the observed 3.6 $\mu$m - CXB cross-correlation would remain unexplained.

Given the results on the DGL obtained so far, it is important to re-examine the DCBH scenario in \citet{2013MNRAS.433.1556Y} to  
assess whether DCBHs can simultaneously provide the extra power at intermediate scales \textit{and} the observed 3.6 $\mu$m-CXB cross-correlation. DCBH forms out of pristine gas within atomic-cooling halos ($T_{\rm vir} \ge 10^4$~K) that are irradiated by a strong external flux in the Lyman-Werner or near-infrared bands ($J_{\rm LW}\gsim10^{1-5}\times10^{-21}$ erg s$^{-1}$cm$^{-2}$Hz$^{-1}$sr$^{-1}$, see e.g. \citealt{2010MNRAS.402.1249S}) . In such halos the Ly$\alpha$ transition is the only efficient cooling mechanism, since H$_2$ formation is suppressed. As a result the gas contracts almost isothermally and eventually directly collapses into a central black hole seed of mass $\sim10^{4-6} ~M_\odot$. The seed continues to accrete the surrounding gas, and radiates energy from radio to X-ray with a spectrum derived by, e.g., \citet{2015MNRAS.454.3771P}. 

We will consider two cases: (i) the fiducial model described in \citet{2013MNRAS.433.1556Y}, with a mass density in active DCBHs $\rho_\bullet\approx 4\times10^5~M_\odot$Mpc$^{-3}$ at peak, and (ii) a reduced model in which the  abundance is about 2/3 of the fiducial one, $\rho_\bullet\approx 2.7\times10^5~M_\odot$Mpc$^{-3}$. Note that the DCBH mass density $\sim M_{\rm BH}\times n_{\rm BH}$, so there are degeneracies between the $M_{\rm BH}$ and $n_{\rm BH}$. Here, we actually only vary the $M_{\rm BH}$, while keeping other parameters as in the fiducial model. Results are shown in the bottom panels of Fig. \ref{auto36}. It is clear that the DCBH fiducial model can provide the required power at intermediate scales to and, when added to the DGL and low-$z$ galaxy contributions (solid line), it gives a good fit to {\tt  Spitzer} data at all scales. Moreover, given the uncertainties in the DGL signal determination, even the reduced DCBH model can be accommodated.
 
\begin{figure}
\centering{
\includegraphics[scale=0.4]{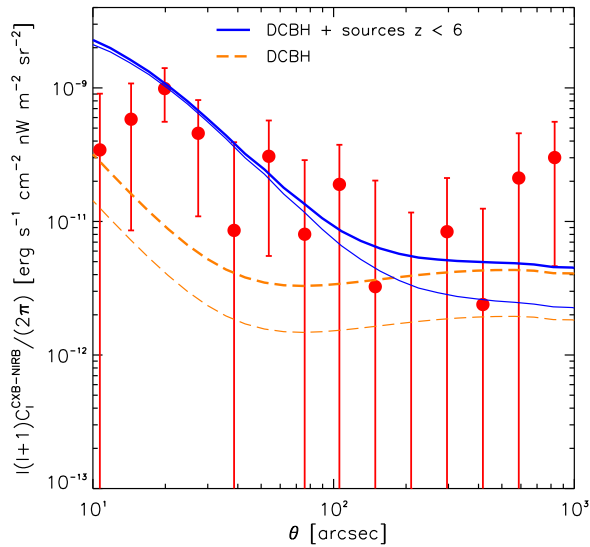}
\caption{$(0.5-2.0)$ keV CXB -- 3.6 $\mu$m IR cross-correlation power spectrum. Points are observations from \citet{2013ApJ...769...68C}; dashed curves show the contribution from DCBHs; solid curves are the sum of DCBH and remaining $z<6$ sources (AGNs, galaxies and hot gas, from \citealt{2014ApJ...785...38H}).  Thick (thin) lines are for the fiducial (reduced) model with $\rho_\bullet = 4\times10^5~M_\odot$Mpc$^{-3}$ ($\rho_\bullet= 2.7\times10^5~M_\odot$Mpc$^{-3}$) at peaks.}
\label{IRX}
}
\end{figure}

We then turn to the CXB. Fig. \ref{IRX} reports the cross-correlation spectrum between CXB in $(0.5-2.0)$ keV \citep{2013ApJ...769...68C} and 3.6 $\mu$m IR fluctuations. We also plot the expected DCBH contribution for the fiducial (reduced) model assuming a typical X-ray equivalent HI column density of  $N_{\rm H} = 1.3\times10^{25}$~cm$^{-2}$, and the sum of the DCBH + low-$z$ sources (AGNs, galaxies and hot gas, modeled by \citet{2014ApJ...785...38H}). Clearly, DCBHs can provide the extra power at large scales required by the data (see \citealt{2013MNRAS.433.1556Y} for more details). We also find that the predicted CXB auto-correlation power spectrum of DCBHs for both the fiducial and reduced model is still well below the measured level \citep{2013ApJ...769...68C}, and does not exceed the unresolved fraction of the CXB intensity at 1.5 keV measured by \citet{2012A&A...548A..87M}. 
\section{Conclusions}
The DCBH scenario accounting for the observed ``clustering excess'' over known galaxies signal in the NIRB power spectrum has been questioned by the recent detection of a correlation between the two  {\tt CIBER} 1.1/1.6 $\mu$m bands with the 3.6 $\mu$m  {\tt  Spitzer} one. This correlation is hardly explained by early DCBHs that, due to intergalactic absorption, cannot contribute to the shortest wavelength bands. We have shown that the new correlation is caused instead by a Diffuse Galactic Light (DGL) component arising from Galactic stellar light scattered by dust. In particular, we have found that:
\begin{itemize}
\item The (1.1)1.6$\times$3.6 $\mu$m cross-correlation power spectra can be fitted nicely by a DGL component that dominates the large scale NIRB fluctuations, and a shot noise component from the remaining low-$z$ galaxies accounting for the small scale power.

\item By assuming perfect correlations (unity correlation coefficient) at different wavelengths, the derived best-fitting DGL fluctuations dominate the 3.6 $\mu$m auto-correlation power spectra on scales $l\lsim 300$. Between $300\lsim l \lsim2\times10^4$ extra sources in addition to the DGL and the remaining low-$z$ galaxies are required to interpret the ``clustering excess".

\item If the 3.6 $\mu$m NIRB fluctuations are only from DGL + remaining low-$z$ galaxies, the observed 3.6-CXB cross-correlation is hard to explain. Introducing the DCBH gives a much better fit of 3.6 $\mu$m auto-correlation power spectra, and naturally explains the 3.6-CXB cross-correlation. It also predicts the much decreased clustering term in NIRB-CXB correlations at $\lsim 1.6$~$\mu$m.

\item Finally, we point out that the DGL inclusion allows to decrease by up to about 30\% the required DCBH abundance/mass. 
\end{itemize}
We conclude that the DCBH scenario remains a viable interpretation of the puzzling NIRB clustering excess, making the NIRB a superb tool to investigate the pristine cosmic epochs in which the supermassive black hole seeds formed.

\section*{ACKNOWLEDGMENTS}

We thank the Euclid-LIBRAE Team for insightful discussions. We are also indebted to M. Zemcov and A. Cooray for help with the interpretation of the {\tt CIBER} data, to A. Cooray and K. Mitchell-Wynne for useful comments.


\begin{thebibliography}{22}
\expandafter\ifx\csname natexlab\endcsname\relax\def\natexlab#1{#1}\fi

\bibitem[{{Cappelluti} {et~al}\mbox{.}(2013){Cappelluti}, {Kashlinsky},
  {Arendt}, {Comastri}, {Fazio}, {Finoguenov}, {Hasinger}, {Mather}, {Miyaji},
  \& {Moseley}}]{2013ApJ...769...68C}
{Cappelluti} N. {et~al.}, 2013, \apj, 769, 68

\bibitem[{{Comastri} {et~al}\mbox{.}(2015){Comastri}, {Gilli}, {Marconi},
  {Risaliti}, \& {Salvati}}]{2015A&A...574L..10C}
{Comastri} A., {Gilli} R., {Marconi} A., {Risaliti} G., {Salvati} M., 2015,
  \aap, 574, L10

\bibitem[{{Cooray} {et~al}\mbox{.}(2012{\natexlab{a}}){Cooray}, {Gong},
  {Smidt}, \& {Santos}}]{2012ApJ...756...92C}
{Cooray} A., {Gong} Y., {Smidt} J., {Santos} M.~G., 2012{\natexlab{a}}, \apj,
  756, 92

\bibitem[{{Cooray} {et~al}\mbox{.}(2012{\natexlab{b}}){Cooray}, {Smidt}, {de
  Bernardis}, {Gong}, {Stern}, {Ashby}, {Eisenhardt}, {Frazer}, {Gonzalez},
  {Kochanek}, {Koz{\l}owski}, \& {Wright}}]{2012Natur.490..514C}
{Cooray} A. {et~al.}, 2012{\natexlab{b}}, \nat, 490, 514

\bibitem[{{Ferrara} {et~al}\mbox{.}(2014){Ferrara}, {Salvadori}, {Yue}, \&
  {Schleicher}}]{2014MNRAS.443.2410F}
{Ferrara} A., {Salvadori} S., {Yue} B., {Schleicher} D., 2014, \mnras, 443,
  2410

\bibitem[{{Helgason} {et~al}\mbox{.}(2014){Helgason}, {Cappelluti}, {Hasinger},
  {Kashlinsky}, \& {Ricotti}}]{2014ApJ...785...38H}
{Helgason} K., {Cappelluti} N., {Hasinger} G., {Kashlinsky} A., {Ricotti} M.,
  2014, \apj, 785, 38

\bibitem[{{Helgason} {et~al}\mbox{.}(2012){Helgason}, {Ricotti}, \&
  {Kashlinsky}}]{2012ApJ...752..113H}
{Helgason} K., {Ricotti} M., {Kashlinsky} A., 2012, \apj, 752, 113

\bibitem[{{Kashlinsky} {et~al}\mbox{.}(2012){Kashlinsky}, {Arendt}, {Ashby},
  {Fazio}, {Mather}, \& {Moseley}}]{2012ApJ...753...63K}
{Kashlinsky} A., {Arendt} R.~G., {Ashby} M.~L.~N., {Fazio} G.~G., {Mather} J.,
  {Moseley} S.~H., 2012, \apj, 753, 63

\bibitem[{{Kashlinsky} {et~al}\mbox{.}(2005){Kashlinsky}, {Arendt}, {Mather},
  \& {Moseley}}]{2005Natur.438...45K}
{Kashlinsky} A., {Arendt} R.~G., {Mather} J., {Moseley} S.~H., 2005, \nat, 438,
  45

\bibitem[{{Kashlinsky} {et~al}\mbox{.}(2007){Kashlinsky}, {Arendt}, {Mather},
  \& {Moseley}}]{2007ApJ...654L...5K}
{Kashlinsky} A., {Arendt} R.~G., {Mather} J., {Moseley} S.~H., 2007, \apjl,
  654, L5

\bibitem[{{Kormendy} \& {Ho}(2013)}]{2013ARA&A..51..511K}
{Kormendy} J., {Ho} L.~C., 2013, \araa, 51, 511

\bibitem[{{Matsumoto} {et~al}\mbox{.}(2011){Matsumoto}, {Seo}, {Jeong}, {Lee},
  {Matsuura}, {Matsuhara}, {Oyabu}, {Pyo}, \& {Wada}}]{2011ApJ...742..124M}
{Matsumoto} T. {et~al.}, 2011, \apj, 742, 124

\bibitem[{{Mitchell-Wynne} {et~al}\mbox{.}(2015){Mitchell-Wynne}, {Cooray},
  {Gong}, {Ashby}, {Dolch}, {Ferguson}, {Finkelstein}, {Grogin}, {Kocevski},
  {Koekemoer}, {Primack}, \& {Smidt}}]{2015NatCo...6E7945M}
{Mitchell-Wynne} K. {et~al.}, 2015, Nature Communications, 6, 7945

\bibitem[{{Moretti} {et~al}\mbox{.}(2012){Moretti}, {Vattakunnel}, {Tozzi},
  {Salvaterra}, {Severgnini}, {Fugazza}, {Haardt}, \&
  {Gilli}}]{2012A&A...548A..87M}
{Moretti} A., {Vattakunnel} S., {Tozzi} P., {Salvaterra} R., {Severgnini} P.,
  {Fugazza} D., {Haardt} F., {Gilli} R., 2012, \aap, 548, A87

\bibitem[{{Pacucci} {et~al}\mbox{.}(2015){Pacucci}, {Ferrara}, {Volonteri}, \&
  {Dubus}}]{2015MNRAS.454.3771P}
{Pacucci} F., {Ferrara} A., {Volonteri} M., {Dubus} G., 2015, \mnras, 454, 3771

\bibitem[{{Seo} {et~al}\mbox{.}(2015){Seo}, {Lee}, {Matsumoto}, {Jeong}, {Lee},
  \& {Pyo}}]{2015ApJ...807..140S}
{Seo} H.~J., {Lee} H.~M., {Matsumoto} T., {Jeong} W.-S., {Lee} M.~G., {Pyo} J.,
  2015, \apj, 807, 140

\bibitem[{{Shang} {et~al}\mbox{.}(2010){Shang}, {Bryan}, \&
  {Haiman}}]{2010MNRAS.402.1249S}
{Shang} C., {Bryan} G.~L., {Haiman} Z., 2010, \mnras, 402, 1249

\bibitem[{{Volonteri} \& {Bellovary}(2011)}]{2011RvMA...23..189V}
{Volonteri} M., {Bellovary} J., 2011, in Reviews in Modern Astronomy, Vol.~23,
  Reviews in Modern Astronomy, {von Berlepsch} R., ed., p. 189

\bibitem[{{Yue} {et~al}\mbox{.}(2013{\natexlab{a}}){Yue}, {Ferrara},
  {Salvaterra}, \& {Chen}}]{2013MNRAS.431..383Y}
{Yue} B., {Ferrara} A., {Salvaterra} R., {Chen} X., 2013{\natexlab{a}}, \mnras,
  431, 383

\bibitem[{{Yue} {et~al}\mbox{.}(2013{\natexlab{b}}){Yue}, {Ferrara},
  {Salvaterra}, {Xu}, \& {Chen}}]{2013MNRAS.433.1556Y}
{Yue} B., {Ferrara} A., {Salvaterra} R., {Xu} Y., {Chen} X.,
  2013{\natexlab{b}}, \mnras, 433, 1556

\bibitem[{{Yue} {et~al}\mbox{.}(2014){Yue}, {Ferrara}, {Salvaterra}, {Xu}, \&
  {Chen}}]{2014MNRAS.440.1263Y}
{Yue} B., {Ferrara} A., {Salvaterra} R., {Xu} Y., {Chen} X., 2014, \mnras, 440,
  1263

\bibitem[{{Zemcov} {et~al}\mbox{.}(2014){Zemcov}, {Smidt}, {Arai}, {Bock},
  {Cooray}, {Gong}, {Kim}, {Korngut}, {Lam}, {Lee}, {Matsumoto}, {Matsuura},
  {Nam}, {Roudier}, {Tsumura}, \& {Wada}}]{2014Sci...346..732Z}
{Zemcov} M. {et~al.}, 2014, Science, 346, 732

\end{thebibliography}
 \end{document}